\documentclass[aps,prl,reprint,showpacs]{revtex4-1}
\usepackage{graphicx}
\usepackage{amssymb}
\usepackage{amsmath}
\usepackage{bm}
\usepackage{verbatim}
\usepackage[colorlinks=true,urlcolor=blue,citecolor=blue,linkcolor=blue]{hyperref}

\newcommand{\addMR}[1]{{\color{black}{#1}}}

\renewcommand{\vec}[1]{{\bm #1}}

\begin{document}

\title{Spin-lattice order in one-dimensional conductors: beyond the RKKY effect}

\author{Michael Schecter, Mark S. Rudner, and Karsten Flensberg}
\affiliation{Center for Quantum Devices and Niels Bohr International Academy, Niels Bohr Institute, University of Copenhagen, 2100 Copenhagen, Denmark}

\date{\today}

\begin{abstract}
We investigate magnetic order in a lattice of classical spins coupled to an isotropic gas of one-dimensional (1d) conduction electrons via local exchange interactions. The frequently discussed Ruderman-Kittel-Kasuya-Yosida (RKKY) effective exchange model for this system predicts that spiral order is always preferred. Here we consider the problem nonperturbatively, and find that such order vanishes above a critical value of the exchange coupling that depends strongly on the lattice spacing. The critical coupling tends to zero as the lattice spacing becomes commensurate with the Fermi wave vector, signalling the breakdown of the perturbative RKKY picture, and spiral order, even at weak coupling. We provide the exact phase diagram for arbitrary exchange coupling and lattice spacing, and discuss its stability. Our results shed new light on the problem of   utilizing a spiral spin-lattice state to drive a one-dimensional superconductor into a topological phase.
\end{abstract}


\maketitle

\emph{Introduction} $\--$  The RKKY coupling mechanism serves as a cornerstone of our understanding of the indirect exchange interaction between magnetic impurities in a metallic host \cite{RK,Kasuya,Yosida}. Recently, the RKKY effect  has played a central role in the intense effort to realize topological superconductivity and associated Majorana bound states in systems of magnetic impurities placed on $s$-wave superconductors \cite{Braunecker-Simon,Yazdani-Loss,Vazifeh-Franz,Franz-2014,VO1,VO2,VO3}.
In a similar setup, promising evidence for Majorana states was recently reported \cite{Yazdani1,Yazdani2} from experiments \addMR{in which a scanning tunneling microscope was used} to create and probe a chain of ferromagnetically aligned Fe atoms on the surface of bulk superconducting Pb. Although the physics of this specific realization is likely dominated by direct Fe-Fe exchange interactions, spin-orbit coupling and magnetic anisotropy, \addMR{these exciting results provide motivation to develop a}  
deeper understanding of possible ``self-organized'' phases of magnetic impurities embedded in electronic environments.

 For \addMR{a one-dimensional (1d) system without spin-orbit coupling}, 
the emergence of topological superconductivity is directly tied to the presence of spiral order in the magnetic impurity chain.
\addMR{Such ordering has been predicted to} arise naturally in the system's ground state due to the RKKY coupling between magnetic impurities \cite{Braunecker-Simon,Yazdani-Loss,Vazifeh-Franz,Franz-2014}. 
Focusing on the normal state of the host, the RKKY coupling arises 
\addMR{at second order} in the direct exchange between conduction electrons and impurities. 
Because the electron wavefunctions oscillate in space with the Fermi wavevector $k_F$, the mediated interaction exhibits characteristic $2k_{\mathrm{F}}$ oscillations: $H_{\mathrm{RKKY}}\propto -\textbf{S}_1\cdot\textbf{S}_2\,\mathrm{cos}(2k_F r_{12})/r_{12}$, where $\textbf{S}_{1,2}$ represent the spin vectors of magnetic impurities separated by a distance $r_{12}$. The long-range $\sim 1/r$ scaling of the interaction results from the gapless nature of excitations near the \addMR{Fermi points of the 1d host system,} 
and is responsible for the logarithmically diverging static spin susceptibility $\chi(k)$ (Fourier transform of $H_{\mathrm{RKKY}}$) at wavevector $k=2k_F$.
\begin{figure}[t]
\includegraphics[width=\columnwidth]{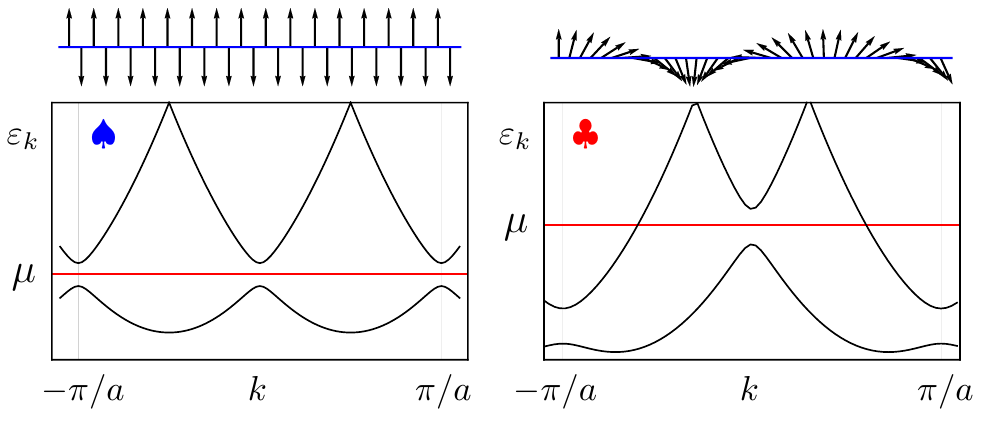}
\caption{(Color online) Electronic band structure for the spin lattice in the antiferromagnetic state (left panel) and spiral phase (right panel). The corresponding points in the phase diagram are shown by the symbols {\color{blue}{$\spadesuit$}} and {\color{red}{$\clubsuit$}} in Fig.~\ref{fig:phase-diagram}. Close to the commensurate points, $k_Fa=(2n+1)\pi/2$, the AF state fully gaps the electronic system. When the lattice spacing is increased the gaps in the AF state become smaller and at a critical value of the lattice spacing the partially gapped spiral state becomes the lowest energy state.}
\label{fig:band}
\vspace{-0.2 in}
\end{figure}

Considering a classical spin lattice subject to this interaction, one easily sees that it lowers its energy $E$ by forming a spiral state with $\textbf{S}_i\cdot\textbf{S}_j=\mathrm{cos}(qr_{ij})$:
\begin{eqnarray}
\nonumber
 E(q)&\propto&-\sum_{i\neq j}\mathrm{cos}(qr_{ij})\mathrm{cos}(2k_Fr_{ij})/r_{ij}
\\
\label{eq:E-RKKY}&=& -\mathrm{Re}\chi(q)\sim \mathrm{log}\left|(2k_F-|q|)a\right|,
\end{eqnarray}
where $a$ is the lattice spacing. Minimizing Eq.~(\ref{eq:E-RKKY}) gives $q=2k_F$, with a logarithmically diverging energy gain \cite{footnote1}.

This \addMR{diverging energy gain} is an artifact of the Born approximation, where renormalization of electron propagation due to spin-lattice scattering \addMR{is neglected}. 
In particular, such scattering exhibits resonant enhancement due to Bragg reflection when $2k_Fa\simeq \pi n$ for integer $n$ and is associated with the opening of electronic band gaps at the Fermi energy. At the odd commensurate points $2k_Fa=(2n+1)\pi$, the system is an antiferromagnet (AF) with a fully gapped electronic bandstructure, while a spiral phase (S) gives a single band gap, see Fig.~\ref{fig:band}. 
\addMR{For any finite coupling, these two bandstructures are not smoothly connected. 
Hence the AF phase may be expected to persist throughout extended intervals of lattice spacings, around the commensurate values.} \addMR{The competition between partially and fully gapped S and AF phases 
demonstrates} the relevance of higher order effective $n$-body spin interactions, neglected in the 2-body spin Hamiltonian $H_{\mathrm{RKKY}}$. 

\begin{figure}[t]
\centering
\includegraphics[width=\columnwidth]{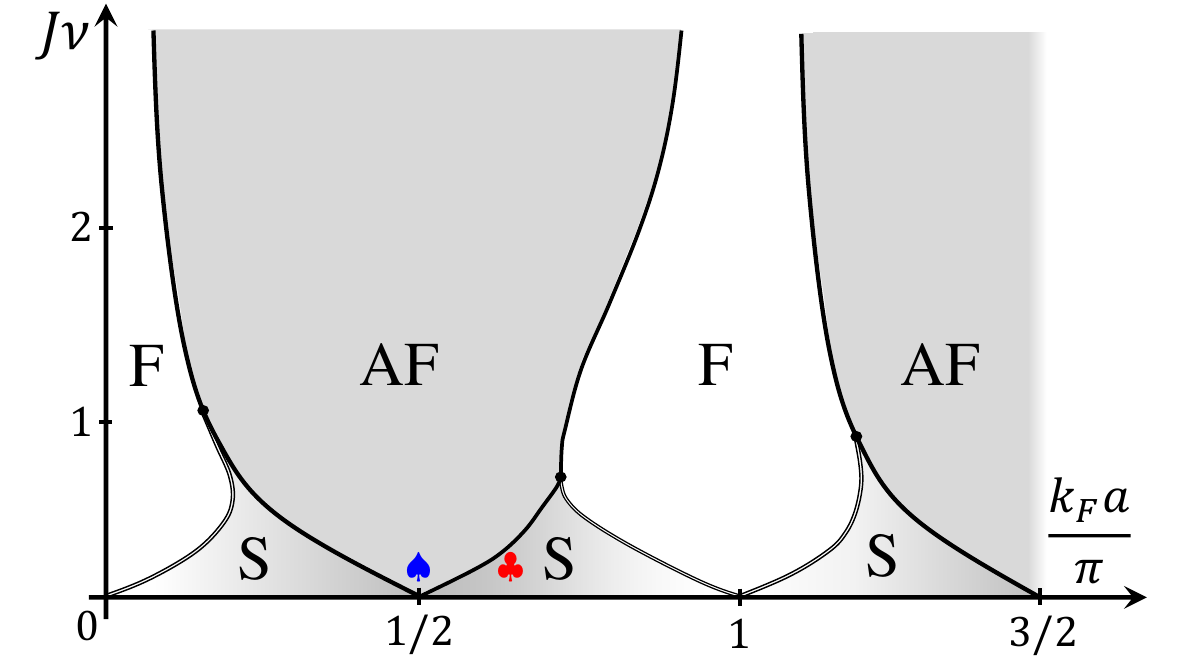}
\caption{(Color online) Groundstate phase diagram of the model in Eq.~(\ref{eq:H}) as a function of lattice spacing $a$ and exchange coupling $J$. Shading schematically represents the wave vector $q$ of the spiral, lying between $q=0$ in the ferromagnetic (F) phase and $q=\pi/a$ in the anti-ferromagnetic (AF) phase. Solid phase boundary lines correspond to first-order transitions, double lines to second-order transitions and dots to triple points. The symbols {\color{blue}{$\spadesuit$}} and {\color{red}{$\clubsuit$}} correspond to the parameters used for the bandstructures shown in Fig.~\ref{fig:band}.}
\label{fig:phase-diagram}
\vspace{-0.2 in}
\end{figure}

In this Letter, we investigate the ground state of a classical spin lattice coupled locally via spin exchange to an isotropic gas of 1d conduction electrons. We find that spin lattice spiral order vanishes above a critical value of the exchange coupling, which depends strongly on the lattice spacing.
\addMR{In particular}, for $2k_Fa\simeq n\pi$ where $n$ is an even (odd) integer, the second (first) order phase transition to the ferromagnetic (anti-ferromagnetic) state occurs even at small exchange coupling. Remarkably, our exact calculation shows that the cumulative effect of all $n$-body interactions (neglected in the RKKY/Born approximation) can drive the system into new phases, even for weak coupling where the RKKY picture is naively expected to be valid, see Fig.~\ref{fig:phase-diagram}.

Similar competing magnetic phases were reported in Refs.~\cite{Shimahara,Pekker,Majumdar}, which focused on the dense impurity limit $k_Fa<1$ in higher dimensions \cite{Shimahara,Majumdar} or in the long wavelength limit $qa\ll1$ \cite{Pekker}. Motivated by the prospect of achieving a topologically non-trivial phase using magnetic adatoms controllably arranged on a superconducting surface, we consider a continuously \addMR{variable} spin-lattice spacing. This, in particular, allows us to address the dilute impurity limit, $k_Fa>1$, and to show that commensurability plays an important role for arbitrary $k_F a$. 

\addMR{In addition to exposing the competing phases,} 
our analysis allows us to determine the exact phase boundaries at arbitrary exchange coupling $J$ and lattice spacing. Interestingly, above a critical exchange coupling $J>J_c \sim1/\nu$ the spiral phase vanishes for \emph{any} lattice spacing (here $\nu=2/(\pi \sqrt{2\mu/m})$, $\mu$ is the chemical potential, \addMR{and $m$ is the effective mass of electrons in the metallic host}). 
\addMR{The existence of the critical coupling implies the appearance of} 
triple points \addMR{in the phase diagram}, where all three phases coexist in thermodynamic equilibrium, see Fig.~\ref{fig:phase-diagram}.

\emph{Model} $\--$ We illustrate the effects described 
above by studying \addMR{the ground states of an electronic system with} the 
Hamiltonian $(\hbar=1)$
\begin{eqnarray}
\label{eq:H}
H=\int dx \left[\sum_\sigma\frac{\psi^\dagger_\sigma \hat{p}^2 \psi_\sigma}{2m}+J_{\mathrm{ex}}\mathbf{S}(x)\cdot\hat{\mathbf{s}}(x)\right],
\end{eqnarray}
where $\psi_\sigma\,(\psi^\dagger_\sigma)$ are real space fermionic annihilation (creation) operators with spin projection $\sigma$, $\hat{p}=-i\partial_x$, $\hat{s}=\frac{1}{2}\psi^\dagger_\sigma\vec{\sigma}_{\sigma\sigma^\prime}\psi_{\sigma^\prime}$ is the electron spin density ($\vec{\sigma}$ is the vector of Pauli matrices) and $J_\mathrm{ex}$ is the exchange interaction constant.
\addMR{We model the magnetic lattice by classical spins, described by} $\textbf{S}(x)=S\sum_j\delta(x-ja)\,\textbf{n}_j$, 
\addMR{where $a$ is the spacing between impurity spins and $\vec{n}_j$ is a unit vector} denoting the direction of the $j^{\mathrm{th}}$ spin.
The quantum limit of a Kondo lattice of spins with $S=1/2$ has a separate, rich history \cite{Ueda1,Honner1,Honner2,Ueda2,Avignon,Rosengren}.

\addMR{To investigate magnetic ordering in the system described by Hamiltonian (\ref{eq:H}), we use a variational approach in which we minimize the electronic ground state energy with respect to the magnetic ordering profile.}
We consider a planar spiral order ansatz of the form $\mathbf{n}_j=\left(\mathrm{cos}\,q a j,\mathrm{sin}\,q a j,0\right)$ which interpolates between ferromagnetic (F) $(q=0)$ and anti-ferromagnetic (AF) $(q=\pm\pi/a)$ order \cite{footnote2}. 

\addMR{To evaluate} the groundstate energy of Hamiltonian (\ref{eq:H}), 
\addMR{we first remove} the winding of the ordering field using the unitary transformation  $U=\mathrm{e}^{i qx \sigma_z/2}$ \cite{DasSarma-RKKY,Braunecker-Simon,Yazdani-Loss,Vazifeh-Franz,Franz-2014,Braunecker2010}. The transformed Hamiltonian $H^\prime=UHU^\dagger$ 
\addMR{has} discrete translational symmetry, 
\addMR{as well as} an effective spin-orbit coupling in the direction perpendicular to the spiral plane,

\begin{equation}
\label{eq:H1}
H^\prime=\int_x \left[\sum_\sigma\frac{\psi^\dagger_\sigma\left(\hat{p}-\frac{1}{2}q\sigma_z\right)^2\psi_\sigma}{2m}+J_{\mathrm{ex}}\mathbf{S}^\prime(x)\cdot\hat{\mathbf{s}}(x)\right],
\end{equation}
where $\textbf{S}^\prime=S\sum_j\delta(x-ja)\hat{x}$ is related to $\textbf{S}$ by an SO(3) rotation around the $z$-axis. The Hamiltonian $H^\prime$ is reminiscient of models of quantum wires with intrinsic spin-orbit coupling in a homogeneous perpendicular magnetic field \cite{Lutchyn,Oreg,Alicea-review,Leijnse-Flensberg}.
However, 
the discrete and self-ordering nature of the effective magnetic field produced by the spin-lattice \addMR{are essential in our work, and lead to interesting new phenomena.}

\addMR{Within the variational approach, we determine} the \addMR{optimal} wave vector $q$ of the spin lattice 
by minimizing the groundstate free energy $E_0(q)=\langle H-\mu N\rangle_0$ for fixed $\mu$ ($N$ is the total electron number). Up to a $q$-independent constant, $E_0(q)$ can be written in terms of the shift in the electronic density of states, $\delta\nu$, due to the presence of the spin-lattice potential: $E_0(q)=\int^\mu_{-\infty}d\varepsilon\,(\varepsilon-\mu)\delta\nu(\varepsilon)=\int^0_{-\infty}d\varepsilon\,\varepsilon\delta\nu(\varepsilon+\mu)$.     Using Lloyd's formula \cite{Lloyd1,Lloyd2}, we express the shift in the density of states in terms of the free electron Green's function $G_0$ and the spin-dependent potential $V$: $\delta\nu(\varepsilon)=-\frac{1}{\pi}\mathrm{Im\,Tr}\,\partial_\varepsilon\mathrm{ln}(1-G_0(\varepsilon)V)$.

\addMR{In order to evaluate the shift of the density of states, $\delta\nu(\varepsilon)$, we exploit} 
the discrete translational symmetry 
of $H^\prime$, which allows it to be block diagonalized 
using a set of Bloch states labeled by quasi momentum $|k|<\pi/a$ (modulo $2\pi/a$). 
\addMR{Importantly,} the delta function lattice acts as a rank-1 operator within the subspace of states \addMR{for each value of the} quasi momentum $k$.
\addMR{Using the Woodbury matrix identity, 
we can thus} reduce the operator trace occuring in $\delta\nu$ to an integral over the Brillouin zone (the remaining trace over the $2\times2$ spin subspace can be computed explicitly). 
\addMR{In this way we obtain the ground state energy  (per system length $L$) for each fixed $q$, measured relative to the value with $J = 0$:} 
\begin{equation}
\label{eq:E}
\frac{E_0(q)}{L}=-\frac{1}{\pi}\mathrm{Im}\int\limits^0_{-\infty}\!\!d\varepsilon\,\varepsilon\!\!\!\int\limits_{-\pi/a}^{\pi/a}\frac{dk}{2\pi}\partial_{\varepsilon}\mathrm{ln}\left(1-\frac{J^2}{a^2}\mathcal{G}^+_0\mathcal{G}^-_0\right),\!\!
\end{equation}
where $\mathcal{G}^\pm_0(\varepsilon,k)=\sum_{m\in\mathbb{Z}} G_0^{\pm}(\varepsilon,k+2\pi m/a)=(ma/\alpha)\mathrm{sin}\alpha a\left[\mathrm{cos}\left(ka\mp qa/2\right)-\mathrm{cos}\alpha \right]^{-1}$ is the \addMR{Green function for free electrons (i.e.,~evaluated for $J = 0$) with conserved spin projection $\pm 1$ along the $z$-axis,}  
and $\alpha=\sqrt{2m(\varepsilon+i0^++\mu)}$.
\begin{figure}[t]
\centering
\includegraphics[width=1.1\columnwidth]{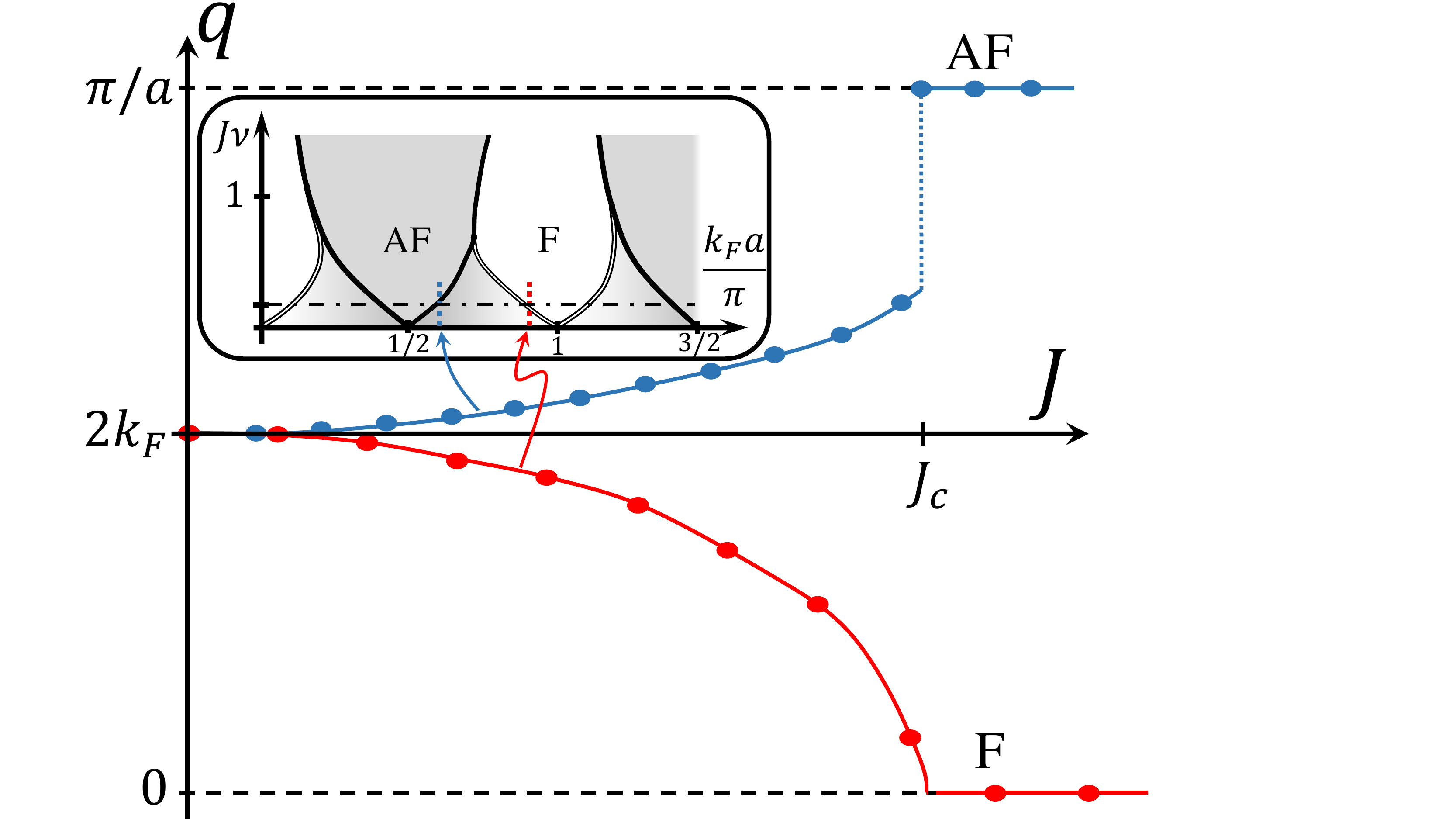} 

\caption{(Color online) 
Optimal wave vector of the spin spiral $q$ as a function of exchange coupling at $k_Fa=0.9\pi$ (red) and $k_Fa=0.6\pi$ (blue), shown as vertical cuts of the inset.}
\label{fig:q-opt1}
\end{figure}
Minimization of $E_0(q)$ gives $q=q_{\mathrm{optimal}}$ which we denote as $q$ for brevity. 
We associate $q=0$ with F, $q=\pi/a$ with AF and intermediate values with S, see Fig.~\ref{fig:phase-diagram}.

As noted in Refs.~\cite{Braunecker-Simon,Yazdani-Loss,Vazifeh-Franz,Franz-2014}, the physics underlying the ordering of the spin-lattice can also be understood by analyzing the corresponding electronic band structure. From that point of view, the advantage of the $q=2k_F$ spiral 
comes from 
the opening of a gap for half the degrees of freedom at the Fermi surface, see Fig.~\ref{fig:band}, \addMR{which lowers the energies of occupied states.} 
This ``spin-Peierls effect" is responsible for the existence of the stable spiral phase, but ultimately the influence of resonant Bragg reflection leads to a transition near points of commensurability.

\begin{figure}[t]
\centering
\includegraphics[width=1.1\columnwidth]{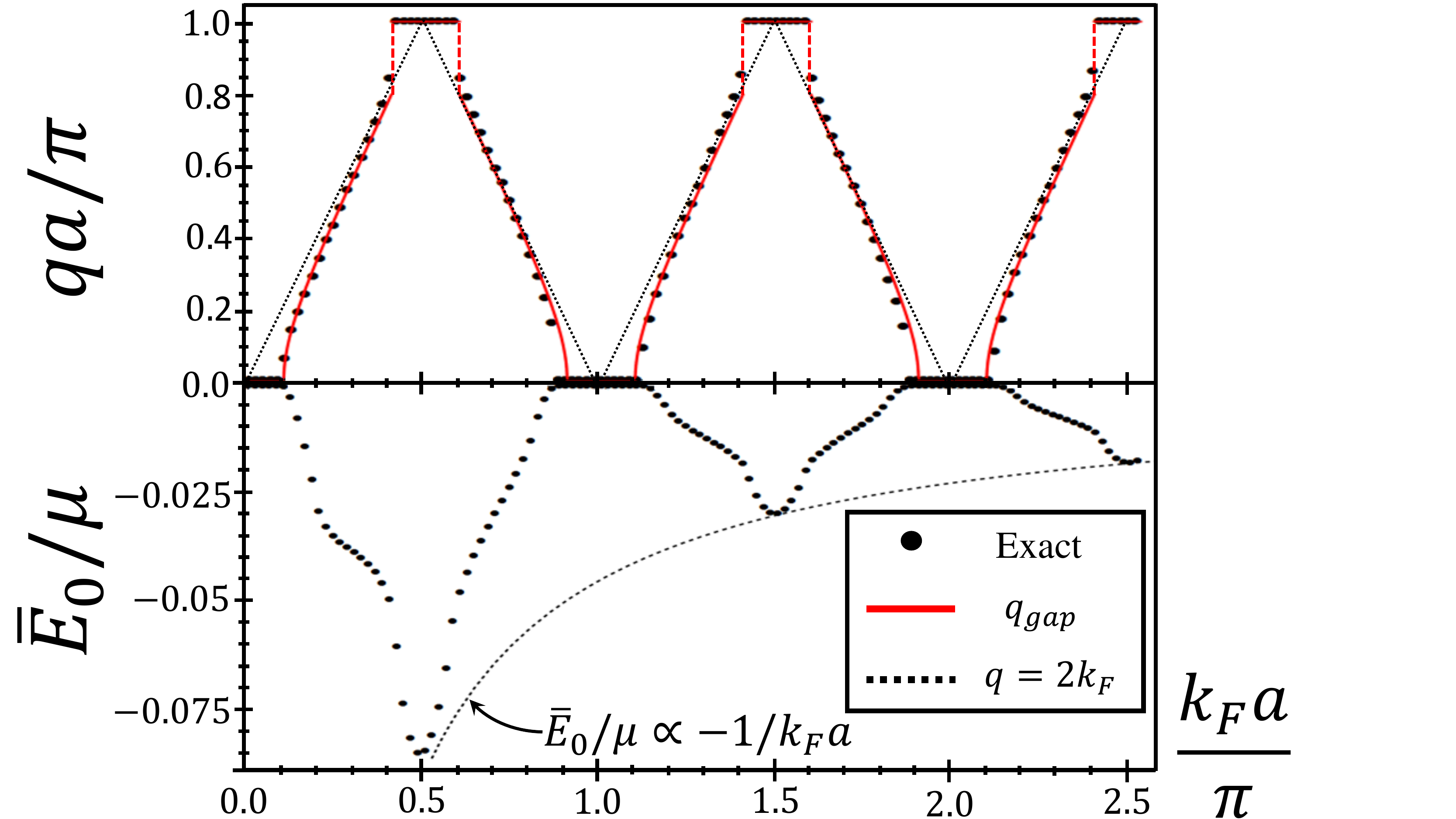} 
\caption{(Color online) \addMR{Optimal} wavevector $q$ and energy per impurity \addMR{spin} $\bar{E}_0=E_0a/L$ as a function of $k_Fa$ at $J\nu=0.2$, shown as horizontal cut of the inset in Fig.~\ref{fig:q-opt1}. The energy is measured relative to the $q=0$ state.}
\label{fig:q-opt2}
\end{figure}

\emph{Phase diagram for $J\nu\ll1$} $\--$ The pole structure of Eq.~(\ref{eq:E}), and thus the electronic band structure, is determined by the \addMR{zeros of the argument of $\mathrm{ln}(1-\frac{J^2}{a^2}\mathcal{G}^+_0\mathcal{G}^-_0)$. Substituting in the expressions for $\mathcal{G}^\pm_0$ gives} 
\begin{equation}
\label{eq:spectral}
1=\frac{\frac{m^2 J^2}{\alpha^2}\sin^2\alpha a}{\left(\cos k_+a- \cos \alpha a\right)\left(\cos k_-a-\cos \alpha a\right)},
\end{equation}
where $k_\pm=k\pm q/2$. In order for the spin-Peierls effect to \addMR{yield the maximal} energy gain, the gap which it opens should be centered at the Fermi energy: $\varepsilon=\pm\Delta$. This relatively simple (heuristic) condition on the gap, when imposed on the exact spectral equation (\ref{eq:spectral}), provides an approximation for the wave-vector of the spiral order,
\begin{equation}
\label{eq:q}
q_{\mathrm{gap}}a=2\mathrm{Re}\left[\mathrm{arccos}\left(\sqrt{1+(\pi J\nu/2)^2}\left|\mathrm{cos}k_Fa\right|\right)\right],
\end{equation}
which agrees quite well with the exact numerical result of Eq.~(\ref{eq:E}), see red solid lines in Fig.~\ref{fig:q-opt2}.

\addMR{Remarkably, the approximate relation in} Eq.~(\ref{eq:q}) correctly shows that the RKKY result $q=2k_F$ is recovered only \addMR{in the limit $J\rightarrow 0$.} 
\addMR{More generally,} Eq.~(\ref{eq:q}) predicts a critical value  $J_c^{(\mathrm{F})}=\frac{2}{\pi\nu}\left|\mathrm{tan}k_Fa\right|$ at which \addMR{the optimal spiral wavevector} goes to zero continuously, $q\sim(J_c^{(\mathrm{F})}-J)^{1/2}$, \addMR{signalling} a second order phase transition to the F state, see Fig.~\ref{fig:q-opt1}.
\addMR{In the F state, the} electronic states at the Fermi surface are spin-polarized. Although not exact, the phase boundaries predicted by Eq.~(\ref{eq:q}) provide a good qualitative description of the F-S boundaries in Fig.~\ref{fig:phase-diagram}.

The S state also competes with the AF order near half integer values of $k_{F}a/\pi$.
\addMR{At these points,} the system may become \emph{fully} gapped, potentially lowering its energy even further. The transition roughly takes place when the AF band gaps close, although we generally find that when the AF state is the groundstate the electronic bands are always gapped: the system is a band insulator, see Fig.~\ref{fig:band}. Using Eqs.~(\ref{eq:spectral}$\--$\ref{eq:q}), this gives the 1st-order transition to the AF state at $J>J_c^{(\mathrm{AF})}=\frac{2}{\pi\nu}\left|\mathrm{cot}k_Fa\right|$ and $q_{\mathrm{gap}}^{\mathrm{S-AF}}=\frac{2}{a}\mathrm{arccos}(\pi J\nu/2)\approx\frac{\pi}{a}\left(1-J\nu\right)$. Thus the jump of the order parameter is $\delta q/q=J\nu$, to leading order in $J\nu$.

We thus arrive at the following sketch of the phase diagram for $J\nu\ll1$. For $J<\frac{2}{\pi\nu}\mathrm{min}\left(\left|\mathrm{tan}k_Fa\right|,\,\left|\mathrm{cot}k_Fa\right|\right)$, the ground state wavevector is \addMR{approximately} given by Eq.~(\ref{eq:q}); otherwise, $q=0$ (F) or $\pi/a$ (AF). As a result, $J_c$ becomes arbitrarily small when $2k_Fa\sim\pi n$ for integer $n$, implying that the validity of the RKKY approximation is 
restricted to $J\nu/\left|k_Fa-n\pi/2\right|\ll1$ (not $J\nu\ll1$). This analysis also suggests the presence of triple points for $J\nu\simeq1$ which are verified by the exact results (Fig.~\ref{fig:phase-diagram}).

\emph{Phase diagram for $J\nu\gg1$} $\--$ In this case the phase diagram is dominated by the competition between F and AF configurations. In the limit $k_Fa\ll1$ the dominant contribution to the energy comes from occupation of the deep bound state bands formed by the strong spin-lattice potential. For $k_Fa\gg1$ the bound state bands play little role and the phase diagram is controlled by positive energy bands of the scattering continuum.
After a more involved analysis of the spectral equation (\ref{eq:spectral}), one finds that the F state exists only in narrow regions around points of (even) commensurability such that $J\nu<1/\left|k_Fa-\pi n/2\right|$ for even integer $n$.


\emph{Discussion} $\--$ The structure of the phase diagram in Fig.~\ref{fig:phase-diagram} is stable against a variety of perturbations. Generally, we find that modifications remain small as long as the associated energy scale is small compared to $J/a$. Perturbation theory can then be used to estimate changes in the phase boundary lines.

\addMR{For nonzero} temperature $T$, thermal spin fluctuations tend to deplete the magnetic order, which can be stabilized at $T>0$ either by a finite chain size or a magnetic field/anisotropy, see \cite{Braunecker-Simon,Yazdani-Loss,Vazifeh-Franz,Franz-2014} for a detailed discussion on this point. 
\addMR{Thermal fluctuations also modify} the occupation of electronic states, which changes the phase boundaries in Fig.~\ref{fig:phase-diagram}. The associated increase in free energy is minimized in the (band insulator) AF state, where the dependence is exponential $\delta F(T)\sim e^{-J/aT}$. The free energy increases faster for the gapless S and F states, going as $\nu(E_F)T^2$. The exact density of states can be studied using Eq.~(\ref{eq:spectral}) and shows that for $k_Fa<1$ the F state is favored. Thus, the S region for  $k_Fa<\pi/2,\,J\nu<1$ shrinks with growing $T$.

The phase boundaries are also stable under inclusion of weak electron-electron interactions. In the presence of a local density-density interaction $H_{\mathrm{int}} \propto g\int_x(\sum_\sigma\psi^\dagger_\sigma\psi_\sigma)^2$, one can use first order perturbation theory to determine the shift of the phase boundaries. At $k_F a<1$ the F phase corresponds to a spin-polarized Fermi sea and is thus unaffected to leading order, while the energy in the S phase 
\addMR{shifts by an amount proportional to $g$.} 
The S-phase is thus diminished (enhanced) for $g>0$ ($g<0$). Determining the direction of the shift at the S-AF phase boundary requires further analysis.

More interesting is the role of intrinsic SO coupling of the electron gas. By choosing the plane of the spiral to align with the SO axis \cite{footnote3}, we see that the inclusion of intrinsic SO coupling merely shifts the ordering wave vector to a net value: $q\to q-q_{\mathrm{SO}}$, leaving the phase diagram unaltered \cite{DasSarma-RKKY}. Thus, if the parameters  $(J,a)$ of the system lie in the F phase of Fig.~\ref{fig:phase-diagram}, the groundstate of the spin lattice will have the wave vector $q=q_{\mathrm{SO}}$, which acts to ``screen'' the intrinsic SO coupling and {renders the electronic bands ferromagnetic in the spiral-rotated reference frame}. Similarly, bands in the AF phase also have a well-defined spin projection in the same frame. 

This ``spin-lattice screening effect'' has important implications for the prospect of achieving a one-dimensional topological superconductor and associated Majorana bound states by a proximity induced  $s-$wave pairing\cite{Braunecker-Simon,Yazdani-Loss,Vazifeh-Franz,Franz-2014,VO1,VO2,VO3,kjaergaard2012,Klinovaja2012}. 
\addMR{In particular, note that} 
the topological phase can only occur in the S phase in Fig.~\ref{fig:phase-diagram} (if the pairing is local in space). This is because the spin rotation induced by the net ordering wave vector needs to be non-collinear at the two Fermi points, which is only the case in the S regions of Fig.~\ref{fig:phase-diagram}.
This complication 
may be overcome to some extent by the pinning effect of an easy-axis magnetic anisotropy energy $\sim -(S_z)^2$, which favors collinear order in the direction perpendicular to the substrate surface (an easy-plane merely pins the spiral plane). 
 As such, easy-axis anisotropy competes with the tendency of the spin-lattice to form the planar screening spiral.

The one-dimensional regime can be realized if $\mu,\,T$ are smaller than the transverse level spacing of a quasi one-dimensional wire. Occupation of higher subbands leads to a dimensional crossover. 
\addMR{Here we expect the S phase to be to suppressed}, 
since the $2k_F$ instability in magnetic susceptibility does not exist in higher dimensions for the normal state \cite{footnote1,DasSarma-RKKY}. 
In the superconducting state, however, \addMR{this suppression may not hold} 
due to the presence of resonantly enhanced spin-exchange \addMR{via} Shiba states \cite{Yao-2014}.
\addMR{Understanding the role of such effects} requires further investigation.


\emph{Conclusion} $\--$ The magnetic phase diagram of a spin lattice embedded in a one-dimensional conductor exhibits a rich structure that strikingly reveals physics beyond the RKKY/Born approximation. The simplified model (\ref{eq:H}) admits an exact solution, which allowed us to deduce the precise phase boundaries of the spiral order and determine the limitations of the RKKY approach. Our results clarify the regimes where topological superconductivity based on a ``self-organized" spin-lattice state can be realized.

\emph{Acknowledgements} $\--$ The research was supported by the Danish National Research Council, the Villum Kann Rasmussen Foundation, and by the People Programme (Marie Curie Actions) of the European Union's Seventh Framework Programme (FP7/2007-2013) under REA grant agreement PIIF-GA-2013-627838.


\begin{thebibliography}{99}
\bibitem{footnote1} Already for $d=2$ the RKKY model instead predicts ferromagnetic ($q=0$) or anti-ferromagnetic ($q=\pi/a$) order \cite{DasSarma-RKKY}. This is a result of the slower $\sim1/r^2$ decay of the induced spin-spin exchange interaction and thus an absence of a divergent spin susceptibility at $q=2k_F$.
\bibitem{footnote3} The orientation of the spiral is not fixed by the intrinsic SO coupling since the latter can be eliminated via a spin-dependent gauge transformation, analogous to the relation between Eqs.~(\ref{eq:H}), (\ref{eq:H1}). In the rotated frame, the system exhibits rotational invariance and the plane of the spiral may be chosen arbitrarily.
\end{thebibliography}
\end{document}